\documentclass[pra,twocolumn,showpacs,superscriptaddress,floatfix]{revtex4}
\usepackage{epsfig}
\usepackage{latexsym}
\usepackage{amsmath}
\usepackage{graphics}
\usepackage{bm}

\newcommand\br{\mathbf{r}}

\newcommand\bk{\mathbf{k}}

\newcommand\romand{\mathrm{d}}

\begin{document}
\title{Instabilities and the roton spectrum of a quasi-1D
Bose-Einstein condensed gas with dipole-dipole interactions}
\author{Stefano Giovanazzi}
\affiliation{School of Physics \& Astronomy,
  University of St Andrews, North Haugh, St Andrews KY16 9SS, Scotland}
\author{Duncan H. J. O'Dell}
\affiliation{Dept of Physics \& Astronomy,
    University of Sussex,
     Falmer, Brighton BN1 9QH, England}

\date{\today}

\begin{abstract}
We point out the possibility of having a roton-type excitation
spectrum in a quasi-1D Bose-Einstein condensate with dipole-dipole
interactions. Normally such a system is quite unstable due to the
attractive portion of the dipolar interaction. However, by
reversing the sign of the dipolar interaction using either a
rotating magnetic field or a laser with circular polarization, a
stable cigar-shaped configuration can be achieved whose spectrum
contains a `roton' minimum analogous to that found in helium II.
Dipolar gases also offer the exciting prospect to tune the depth
of this `roton' minimum by directly controlling the interparticle
interaction strength. When the minimum touches the zero-energy
axis the system is once again unstable, possibly to the formation
of a density wave.
\end{abstract}
\maketitle

\section{Introduction}
Ultra-cold gases of atoms which interact with long-range
anisotropic dipole-dipole interactions have been the subject of a
number of theoretical analyses over the past few years
\cite{yi,goral,santos00,lushnikov,giovanazzi2002b,derevianko,yi04,bohn,baranov,odell04}.
The very recent advent of Bose-Einstein condensation of molecules
\cite{greiner} has renewed interest in the investigation of
quantum gases with dipolar interactions since molecules can
potentially possess large dipole moments. One of the novel
properties that has been predicted for gaseous Bose-Einstein
condensates (BECs) with dipole-dipole interactions is a `roton'
minimum in the excitation spectrum \cite{odell03,santos03}, a
feature which is absent in the usual case of repulsive short-range
s-wave interactions, and seems to originate in the long-range and
partially attractive nature of dipolar interactions.

The `roton' minimum can be interpreted with the help of Feynman's
formula for the dispersion relation for excitations of energy $E$
and momentum $\hbar k$ of a Bose liquid \cite{pines+nozieres}
\begin{equation}
E(k)=\frac{\hbar^{2} k^{2}}{2 m S(k)}
\end{equation}
where $m$ is the atomic mass. This remarkable formula relates the
excitation spectrum to the two-particle correlation properties as
encapsulated in the static structure factor $S(k)$, which is the
Fourier transform of the pair correlation function. A peak in
$S(k)$ due to strong two-particle correlations can lead to a
minimum in the energy spectrum. The most famous example of a roton
minimum (and the system for which the name was first coined, by
L.D. Landau) is helium II where the roton minimum occurs at
wavelengths coinciding with the average interparticle separation.
The roton minimum in helium II has often been regarded as
indicating that the superfluid is very close to crystallization
\cite{schneider,pitaevskii84,noziereslecture} since the ordering
of atoms on a crystal lattice would lead to a corresponding peak
in $S(k)$. However, many questions concerning this superfluid to
density-wave transition remain difficult to answer, essentially
because liquid helium is a strongly correlated system. Our aim in
this paper is to explore an analogous `roton' spectrum in a
quasi-1D BEC with dipole-dipole interactions. The tunability of
the various parameters (interaction strength, density, effective
dimensionality etc) controlling these relatively simple quantum
gases  gives the exciting possibility to vary the depth of the
roton minimum and hence explore the onset of this zero-temperature
phase transition.

\section{Dipole-Dipole interactions in a Bose-Einstein condensate}
 The long-range part of the interaction between
two dipoles separated by $\br$, and aligned by an external field
along a unit vector $\hat{\mathbf{e}}$, is given by
\begin{equation}
U_{\mathrm{dd}}(\br)= \frac{C_{\mathrm{dd}}}{4 \pi}\, \hat{{\rm
e}}_{i} \hat{{\rm e}}_{j} \frac{\left(\delta_{i j}- 3 \hat{r}_{i}
\hat{r}_{j}\right)}{r^{3}} . \label{eq:staticdipdip}
\end{equation}
Dipoles induced by an electric field
$\mathbf{E}=E\hat{\mathbf{e}}$  have a coupling $
C_{\mathrm{dd}}=E^{2} \alpha^{2}/\epsilon_{0}$, where $\alpha$ is
the static polarizability, and $\epsilon_{0}$ the permittivity of
free space. For atoms with a magnetic dipole moment $d_{m}$
aligned by a magnetic field $\mathbf{B}=B\hat{\mathbf{e}}$, one
has $C_{\mathrm{dd}}=\mu_{0} d_{m}^{2}$, where $\mu_{0}$ is the
permeability of free space. The isotropic short-range part the
interatomic interactions are modelled by a pseudo-potential,
$U_{\mathrm{s}}(\br)=(4 \pi a_{\mathrm{s}} \hbar^{2}/m)
\delta(\br) \equiv g \delta(\br)$,
 where $g$ incorporates the quantum aspects of
low-energy scattering via the $s$-wave scattering length,
$a_{\mathrm{s}}$. In order to apply Bogoliubov (perturbation)
theory to a BEC one generally needs to use such an effective- or
pseudo-potential rather than the bare interaction
\cite{llstatphys2}. In what follows we shall assume, however, that
the dipole-dipole interaction can be treated within the Born
approximation, which corresponds to the Fourier transform of the
bare potential (\ref{eq:staticdipdip}). The Born approximation has
been shown by Yi and You \cite{yi} to work reasonably well for the
long-range dipole-dipole interaction, but see Derevianko's work
\cite{derevianko} for corrections, which can be substantial
\cite{yi04}.

The basic quantity that then enters the Bogoliubov theory is the
Fourier transform of the total effective interaction,
$\widetilde{U}_{\mathrm{tot}}(\bk)= \int \romand^3 \mathrm{r} \exp
[-\mathrm{i} \bk \cdot \br] \{g \delta(\br)
+U_{\mathrm{dd}}(\br)\} $,
\begin{equation}
\widetilde{U}_{\mathrm{tot}}(\bk)=g\left[1+
\varepsilon_{\mathrm{dd}} \, \hat{{\rm e}}_{i} \hat{{\rm e}}_{j}
\left(3 \hat{k}_{i} \hat{k}_{j} - \delta_{i j} \right)\right]
\label{eq:effective}
\end{equation}
where $\varepsilon_{\mathrm{dd}}$ is a dimensionless measure of
the strength of the dipolar interactions relative to the $s$-wave
scattering
\begin{equation}
\varepsilon_{\mathrm{dd}}  \equiv \frac{C_{\mathrm{dd}}}{3 g} \ .
\label{epsilon}
\end{equation}
The definition includes a factor of 3 because for a homogeneous
system (no confining trap) one finds instabilities when
$\varepsilon_{\mathrm{dd}} \ge 1$
\cite{goral,giovanazzi2002b,odell04}. This can be seen directly
from the Bogoliubov dispersion for phonon-like density
perturbations in a homogeneous dipolar BEC \cite{goral}
\begin{equation}
E_{\mathrm{B}}=\sqrt{\left(\frac{\hbar^{2}k^{2}}{2m}\right)^{2}+
2gn\left\{1+\varepsilon_{\mathrm{dd}} \left(3 \cos^{2} \theta -1
\right) \right\}\frac{\hbar^{2}k^2}{2m}} \label{eq:bogdispersion}
\end{equation}
where $n$ is the density. This dispersion relation
(\ref{eq:bogdispersion}) can become imaginary when
$\varepsilon_{\mathrm{dd}}>1$, indicating an instability.  The
interactions enter the dispersion in the second term under the
square root via the effective potential (\ref{eq:effective}). The
dispersion (\ref{eq:bogdispersion}) relation has an angular
dependence ($\theta$ is the angle between the momentum of the
phonon and the external polarizing field) which further
illustrates the richness of dipolar systems in comparison to the
non-dipolar case.

Santos \emph{et al} \cite{santos03} considered the case of a
quasi-2D \emph{pancake}-shaped BEC, where the dipoles are aligned
(by an external static field) along the symmetry axis of the trap.
They predicted that the analogous expression to
(\ref{eq:bogdispersion}) gives a roton minimum in the excitation
spectrum. In this paper, however, we want to see if a roton
minimum can be achieved in the quasi-1D case where the condensate
is \emph{cigar} shaped, being tightly confined in the radial
($x-y$) plane, say, and very elongated along the axial ($z$)
direction. At first sight this seems unlikely because the
anisotropy of the dipole-dipole interaction means that the
quasi-1D and -2D cases are very different. Two dipoles lying
side-by-side in the radial plane are repulsive, whilst two lying
end-to-end along the axial direction are attractive so, broadly
speaking, a pancake shaped BEC tends to be more stable since the
mean-field energy is more positive (repulsive) than that of a
cigar-shaped BEC. Indeed, an infinite quasi-1D BEC with only
dipole-dipole interactions (no repulsive $s$-wave short range
interactions) is unstable to density perturbations, as we shall
see. Therefore we shall employ a special trick and \emph{reverse}
the sign of the dipole-dipole interaction (see also \cite{giovanazzi2002b}). Two methods for
achieving the reversed dipole-dipole interaction will be reviewed in the next section. With the sign
of the dipole-dipole interactions reversed, two dipoles lying
end-to-end along the axial direction are now repulsive and in this
way an infinite quasi-1D dipolar BEC can be stabilized somewhat
against collapse induced by density perturbations.

In our previous work on the possibility of a roton spectrum in a
BEC \cite{odell03} we took the case of laser-induced dipole-dipole
interactions \cite{thirunamachandran}, which on top of the static
interactions of Equation (\ref{eq:staticdipdip}), also have a very
long-range component ($1/r^{2}$, or even $1/r$, depending on the
polarization direction) arising from retardation effects due to
the finite wavelength of the laser radiation. In principle this is
a very general scheme since one can benefit from the large
polarizability of atoms/molecules close to an electronic
transition. The disadvantage of laser-induced dipole-dipole
interactions, in atomic systems at least, is the unavoidable
presence of spontaneous emission which heats the sample. In
molecular systems one may be able to circumvent spontaneous
emission problems somewhat by using transitions in the infra-red
or even microwave regions where spontaneous emission rates are low
and the long wavelength of the radiation allows one to have many
particles in an `interaction volume' (given by
$\lambda_{\mathrm{laser}}^3$). In \cite{giovanazzi2002a} we
discussed how a BEC with light-induced dipole-dipole interactions
is unstable to the formation of a density wave. The periodicity of
the modulation is determined by the wavelength of the laser light
and in one interpretation can be understood in terms of the
back-scattered light interfering with the incident light to form
an optical grating which the atoms then minimize their energy in
by trying to sit predominantly in the wells. In this paper we are
interested in the case of static dipole-dipole interactions.
Although at first sight there are many similarities between the
two cases, in the light-induced case the optical wavelength
provides the fundamental (classical) scale in the problem whereas
here the parameters we must form a length scale from are
$C_{\mathrm{dd}}$, $n$, $m$ and $\hbar$.

\section{Reversing the sign of the dipole-dipole interaction}

i) In the case of magnetic dipoles the sign of the dipole-dipole
interaction (\ref{eq:staticdipdip}) can be reversed by rapidly
rotating a component of the external magnetic field in the radial
plane \cite{giovanazzi2002b}, similarly to a well known technique from solid state NMR. The
magnetic field should have the form
\begin{equation}
\mathbf{B}(t)=B \left\{\cos \phi \hat{\mathbf{z}}+\sin \phi \left[
\cos \left(\Omega t \right) \hat{\mathbf{x}}+ \sin \left( \Omega t
\right) \hat{\mathbf{y}} \right] \right\}
\end{equation}
which is a combination of a static magnetic field directed along
$z$, and a field rotating at frequency $\Omega$ in the
perpendicular radial plane. The frequency should be such that the
dipoles practically do not move during the time $\Omega^{-1}$, but
their moments adiabatically follow the external field
$\mathbf{B}(t)$.  This corresponds to the condition
$\omega_{\mathrm{Larmor}} \gg \Omega \gg \omega_{r}, \
\omega_{z}$, where $\omega_{r}, \ \omega_{z}$ are the radial and
axial trapping frequencies, respectively. Under these conditions
the particles experience a time-averaged interaction whose form is
exactly the same as Equation (\ref{eq:staticdipdip}) except for
being multiplied by $\left(3 \cos^{2} \phi -1 \right)/2$. By
varying the angle $\phi$ this factor can be continuously changed
between -1/2 to 1, giving the possibility to reverse the sign of
the interaction. At the so-called `magic angle' of
$\phi_{M}=54.7^{\circ}$ the dipolar interaction averages to zero.
This technique can be adapted to the case of heteronuclear molecules polarized by a rotating electric field.
ii) For dipoles induced by the electric field of a very long
wavelength laser propagating along the $z$-direction, then one
obtains a reversed dipolar interaction when the laser is
circularly polarized. In the static limit the laser induced
dipole-dipole interaction gives
\begin{equation}
C_{\mathrm{dd}}=I \alpha^{2}/(c \epsilon_{0}^{2})
\end{equation}
where $I$ is the laser intensity and $c$ is the speed of light.
For a laser with wavevector $\mathbf{q}=\hat{\mathbf{z}}
\omega/c$, then application of the identity $\hat{e}_{i}^{\ast
(\pm)}(\mathbf{q})
\hat{e}_{j}^{(\pm)}(\mathbf{q})=\frac{1}{2}[(\delta_{ij}-\hat{q}_{i}
\hat{q}_{j}) \pm \mathrm{i} \epsilon_{ijk} \hat{q}_{k}]$, where
$+(-)$ corresponds to left (right) circular polarizations, leads
to the interaction
\begin{equation}
U_{\mathrm{dd}}^{\mathrm{circ}}(\br)= - \frac{1}{2} \frac{I
\alpha^{2}(0)}{4 \pi c \epsilon_{0}^{2}}\, \frac{\left(1- 3
\cos^{2} \theta \right)}{r^{3}}  \label{eq:laserdipdip}
\end{equation}
where $\theta$ is angle the interparticle separation vector
$\mathbf{r}$ makes with the $z$-axis. We see that the sign of the
interaction has been reversed and multiplied by one half,
corresponding to the maximal change possible in the magnetic case.
In order to keep the notation as simple as possible, whenever
reversed dipolar interactions are used in the rest of this paper
$C_{\mathrm{dd}}$ will be taken to include the multiplying
pre-factors (such as 1/2).

\section{Quasi-1D ansatz}
The dipole-dipole energy functional of the gas is given by
\begin{eqnarray}
H_{\mathrm{dd}} & = & {1\over 2} \int \int \,
 \romand^{3}\mathrm{r} \, \romand^{3}\mathrm{r}' \, n(\mathbf{r})
 \, U_{\mathrm{dd}}(\br-\br')\,  n(\mathbf{r}') \\
& = & {1\over 2}\int \, {\romand^{3} \mathrm{k} \over (2 \pi)^{3}} \,
\widetilde{n}(\bk) \, \widetilde{U}_{\mathrm{dd}}({\mathbf k}) \,
\widetilde{n}(-\bk)
\end{eqnarray}
where $n(\mathbf{r})$ and $\widetilde{n}(\bk)$ are the number
density and its Fourier transform (F.T.), respectively. The F.T.
of the dipole-dipole potential is given in (\ref{eq:effective}).

Assuming an harmonic radial trapping $V_{trap}=(1/2)m\omega_x^2 (x^2+y^2)$
with $\hbar\omega_x$ larger or equal to the mean-field energy, we make a Gaussian ansatz
for the radial part of the number density profile
\begin{eqnarray}
n(\br) & = & N (\pi w_{x}^{2} )^{-1} n^{z}(z) \exp
\left[-(x^{2}+y^{2})/w_{x}^{2} \right] \\
\widetilde{n}(\bk) &= & N \widetilde{n}^{z}(k_z)
\exp[-(k_{x}^2+k_{y}^2)w_{x}^{2}/4]
\end{eqnarray}
where $w_{x}$ is the characteristic radial width and $N$ is the
total number of atoms in the sample. The axial density profile
$n^{z}(z)$ is normalized to one, and its F.T. is
$\widetilde{n}^{z}(k_z)$. Upon inserting the Gaussian ansatz into
the dipolar energy functional, and taking the polarization
direction of the dipoles to be in the axial ($z$) direction the
$H_\mathrm{dd}$ reduces to
\begin{eqnarray}
H_{\mathrm{dd}} &=& (N^{2}/2) \int \int \, \romand z \, \romand z'
\, n^{z}(z) U^{z}_{\mathrm{dd}}(z-z')
 n^{z}(z') \, \\
 &=& (N^{2}/4 \pi) \int \, \romand k_{z} \, \widetilde{n}^{z}(k_{z})
\widetilde{n}^{z}(-k_{z}) \widetilde{U}^{z}_{\mathrm{dd}}(k_{z})
\end{eqnarray}
which defines $\widetilde{U}^{z}_{\mathrm{dd}}(k_{z})$ as the F.T.
of the effective one-dimensional dipole-dipole potential. One
finds that
\begin{equation}
\widetilde{U}^{z}_{\mathrm{dd}}(k_{z})=\frac{C_\mathrm{dd}}{2 \pi
w_x^{2} } \left( \frac{k_z^{2} w_{x}^2}{2} \exp[k_z^{2} w_x^{2}/2]
\, E_1 [k_z^2 w_{x}^2/2] - \frac{1}{3} \right)
\end{equation}
where $E_1[x]=\int_{x}^{\infty} \romand t \, \exp[-t]/t  $ is the
exponential integral \cite{a+s}. In an analogous fashion, the
effective 1D $s$-wave contact interaction in momentum space is
given by $\widetilde{U}^{z}_{\mathrm{s}}=2 \hbar^{2}
a_{\mathrm{s}}/(m w_{x}^2)$.

\begin{figure}[t]
\begin{center}
\centerline{\epsfig{figure=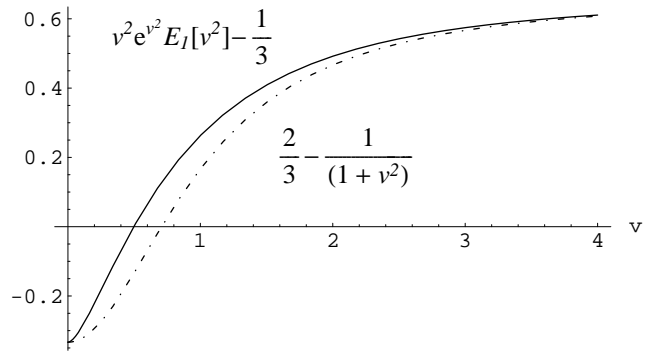, width= 8.5cm}}
\end{center}
 \caption{Solid curve: effective 1D dipolar
potential $\widetilde{U}^{z}_{\mathrm{dd}}(v)$, in momentum space
in units of $C_\mathrm{dd}/(2 \pi w_x^{2})$. Dot-dashed curve: a
simple approximation (as given in Eq.\ (\ref{eq:approx})).}
\label{fig:approxcheck}
\end{figure}

In order to give a qualitative impression of what the effective 1D
potential looks like in coordinate space we note that an
analytically simple approximation to
$\widetilde{U}^{z}_{\mathrm{dd}}$ is given by (see Fig.\
\ref{fig:approxcheck})
\begin{equation}
\widetilde{U}^{z}_{\mathrm{dd}}(k_z) \approx
\frac{C_\mathrm{dd}}{2 \pi w_x^{2}} \, \left(\frac{2}{3} -
\frac{1}{k_z^2 w_{x}^2/2+1} \right). \label{eq:approx}
\end{equation}
Upon transforming back into coordinate space, the $1/(k_z^2
w_{x}^2/2+1)$ term gives an exponential contribution to the
effective coordinate-space 1D potential $U^{z}_{\mathrm{tot}}(z)=
1/(2 \pi) \int \romand k_{z} \exp \left[\mathrm{i} k_{z} z \right]
\widetilde{U}^{z}_{\mathrm{tot}}(k_{z})$, i.e.\
\begin{eqnarray}
U^{z}_{\mathrm{tot}}(z)=\frac{g}{2 \pi w_{x}^2}
\bigg(\left\{2\varepsilon_{\mathrm{dd}}+1\right\}  \delta(z) & \nonumber \\
-  \frac{3\varepsilon_{\mathrm{dd}}}{\sqrt{2} w_{x}} &  \exp
\left[-\sqrt{2}z/w_{r} \right] \bigg).
\end{eqnarray}

\section{Bogoliubov excitation spectrum for a quasi-1D dipolar BEC}
\label{sec:bog} 
\begin{figure}[t]
\begin{center}
\centerline{\epsfig{figure=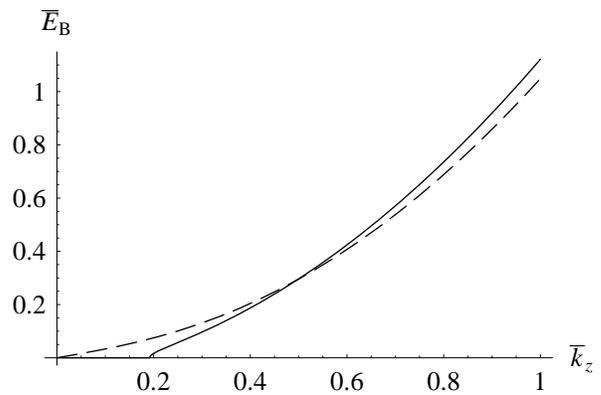, width= 8.5cm}}
\end{center}
\caption{Bogoliubov dispersion relation for axial
excitations in a quasi-1D BEC with $\bar{\mu}=0.1$. Dashed curve:
$s$-wave contact interactions only
($\varepsilon_{\mathrm{dd}}=0$). Solid curve: dipole-dipole plus
$s$-wave contact interactions, $\varepsilon_{\mathrm{dd}}=2$. For
$\bar{k}_{z} < 0.19$ the Bogoliubov energy for the dipolar case is
imaginary. In general, whenever $\varepsilon_{\mathrm{dd}}>1$ the
spectrum is imaginary at long wavelengths, indicating an
instability to density perturbations.} \label{fig:instability}
\end{figure}

The Bogoliubov dispersion relation for excitations
in a homogeneous nearly ideal BEC is
\begin{equation}
E_{\mathrm{B}}^2=c^2p^2+(p^2/2m)^2
\end{equation}
where for a quasi-1D system, with a Gaussian radial profile of
peak density $n(0)$, one has $c^2=\pi n(0) w_x^2
\widetilde{U}^{z}_{\mathrm{tot}}/m$. For the linear parts of the
spectrum $c$ plays the role of the speed of sound. It is useful to
write the Bogoliubov relation in terms of the following
dimensionless quantities
\begin{eqnarray}
\bar{k_z}& = & \frac{k_z w_x}{\sqrt{2}} \\
\bar{\mu} & = & \frac{g n(0)}{\hbar^{2}/(m w_{x}^2)} \\
\bar{E}_{\mathrm{B}} & = & \frac{E_{\mathrm{B}}}{\hbar^{2}/(m
w_{x}^2)}.
\end{eqnarray}
$\bar{k_z}$ is the wavenumber scaled by the inverse radial size,
$w_x$, of the condensate radial wave function ansatz.
The Bogoliubov dispersion relation becomes
\begin{equation}
\bar{E}_{\mathrm{B}}^2=   \bar{\mu} \left(
3\varepsilon_{\mathrm{dd}} \left\{\bar{k}_{z}^{2} \exp [
\bar{k}_{z}^{2} ] E_{1}[\bar{k}_{z}^{2}]-\frac{1}{3}\right\}+1
\right)\bar{k}_z^{2}+ \bar{k}_z^{4}. \label{eq:bogoliubov}
\end{equation}
When $\varepsilon_{\mathrm{dd}}>1$ one finds that the Bogoliubov
energy becomes imaginary for a range of momenta, see Figure
\ref{fig:instability}, indicating an instability due to density
perturbations. If there are no repulsive $s$-wave interactions
($\varepsilon_{\mathrm{dd}} \rightarrow \infty$) to counterbalance
the attractive portion of the dipolar interactions then there is
an instability even for vanishingly small dipole-dipole coupling
strength in this infinite 1D case.
In an actual experiment the trapping potential along the axial
direction can stabilize the BEC somewhat, in an analogous fashion
to the case of a negative $s$-wave scattering length
\cite{bradley97}.

\begin{figure}[t]
\begin{center}
\centerline{\epsfig{figure=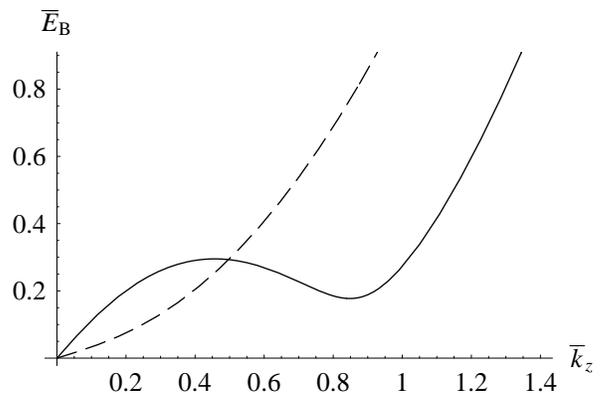, width= 8.5cm}}
\end{center}
\caption{`Roton' minimum in the Bogoliubov dispersion relation for
axial excitations in a quasi-1D dipolar BEC with $\bar{\mu}=0.1$.
Dashed curve: $s$-wave contact interactions only
($\varepsilon_{\mathrm{dd}}=0$). Solid curve: reversed
dipole-dipole plus $s$-wave contact interactions,
$\varepsilon_{\mathrm{dd}}=-13$.} \label{fig:staticroton}
\end{figure}

\begin{figure}[t]
\begin{center}
\centerline{\epsfig{figure=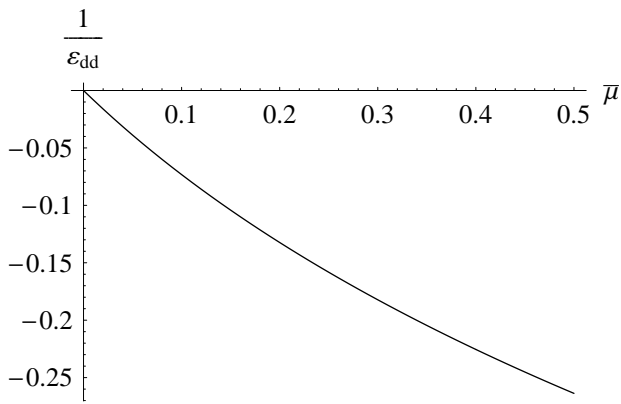, width= 8.5cm}}
\end{center}
\caption{The critical value of
$1/\varepsilon_{\mathrm{dd}}(=3g/C_{\mathrm{dd}})$ when the roton
minimum touches the zero-energy axis plotted as a function of
$\bar{\mu} = \frac{g \, n(0)}{\hbar^{2}/(m w_{x}^2)}$ in an infinite quasi-1D
dipolar BEC.} \label{fig:infinitecyl}
\end{figure}

However, as pointed out earlier, the stability properties of the
condensate can be dramatically changed by reversing the sign of
the dipolar interaction. Figure \ref{fig:staticroton} shows the
Bogoliubov excitation spectrum in the case of negative
$\varepsilon_{\mathrm{dd}}$. In contrast to liquid helium, where
one only has control over macroscopic thermodynamic variables such
as temperature and pressure, in an atomic or molecular BEC
considerable control can be exerted over microscopic quantities
such as the interparticle interactions. A well-known example is
tuning the $s$-wave scattering length using a Feshbach resonance
\cite{inouye}. Similarly, in a system of atoms or molecule with a permanent magnetic or electric dipole, one can
envisage an experiment at a fixed value of $\bar{\mu}$ and
adiabatically changing the dispersion relation by adjusting
$\varepsilon_{\mathrm{dd}}$ via the angle of the rotating external
field. Alternatively, for dipoles induced by an electric field one
can imagine slowly changing the field strength. Within the current
model we find that as $\varepsilon_{\mathrm{dd}}$ is increased in
magnitude below zero (made more negative), first an inflexion
point appears in the dispersion relation and then a fully fledged
roton minimum as shown in Figure \ref{fig:staticroton}. As $\vert
\varepsilon_{\mathrm{dd}} \vert $ is increased yet further the
roton minimum deepens and eventually touches the zero-energy axis,
at which point, following the earlier discussion of helium, we
might expect an instability towards a density-wave. The value of
the wavenumber when the roton touches the zero-energy axis is
found to be
\begin{eqnarray}
(\bar{k}_{z}^{\mathrm{crit}})^{2} & = & -\frac{\bar{\mu}}{2}
\left(2\varepsilon_{\mathrm{dd}}+1 \right) \nonumber \\ & &
-\frac{1}{2}\sqrt{\bar{\mu}^{2}
(2\varepsilon_{\mathrm{dd}}+1)^{2}+4\bar{\mu}(\varepsilon_{\mathrm{dd}}-1)}.
\label{eq:instabcondition1}
\end{eqnarray}
The particular values of the dimensionless chemical potential
$\bar{\mu}$, which is representative of the density, and the
dipolar interaction strength, as represented by
$\varepsilon_{\mathrm{dd}}$, at which this instability occurs obey
the transcendental equation
\begin{equation}
3 \bar{\mu}
\varepsilon_{\mathrm{dd}}(\bar{k}_{z}^{\mathrm{crit}})^{2} \exp[
(\bar{k}_{z}^{\mathrm{crit}})^{2}]
E_{1}[(\bar{k}_{z}^{\mathrm{crit}})^{2}]= \bar{\mu}
(\varepsilon_{\mathrm{dd}}-1)-(\bar{k}_{z}^{\mathrm{crit}})^{2}.
\label{eq:instabcondition2}
\end{equation}
By substituting Equation (\ref{eq:instabcondition1}) into
(\ref{eq:instabcondition2}), one can numerically obtain the
relationship between $\varepsilon_{\mathrm{dd}}$ and $\bar{\mu}$
at the instability. The results are shown in Figure
\ref{fig:infinitecyl}.

 In the case of pure dipolar interactions
(no $s$-wave) the point at which the roton touches the zero-energy
axis can be reduced to a single number
\begin{equation}
\frac{n(0) C_{\mathrm{dd}}}{\hbar^{2}/(m w_{r}^{2})}\approx -3.6.
\label{eq:instabcondition3}
\end{equation}
In this limiting case the wavelength associated with the roton
minimum is given by $\lambda_{\mathrm{roton}}=\sqrt{3}\pi \hbar/
\sqrt{m n(0) C_{\mathrm{dd}}}$, which sets the scale for the
density-wave that we expect to form.

\section{Effect of axial trapping upon the roton instability}
Up till now we have discussed the idealized case of an infinite
quasi-1D cylindrical BEC. We now want to move closer to the
possible experimental situation where the trap is in fact 3D, but
is still assumed to be highly elongated and have cylindrical
symmetry. The ansatz for the condensate density may now be taken
as a 3D Gaussian
\begin{equation}
n(\br)=\frac{N}{\pi^{3/2} w_x^{2} w_{z}} \exp
\left[-(x^{2}+y^{2})/w_{x}^{2}-z^{2}/w_{z}^{2} \right].
\end{equation}
Using this ansatz, the contribution of the dipole-dipole
interactions to the total energy is found to be given by the
expression \cite{yi,giovanazzi2002b,odell04}
\begin{equation}
H_{\mathrm{dd}}=-\frac{N^{2} C_{\mathrm{dd}}}{12 \pi \sqrt{2 \pi}}
\frac{f (\kappa)}{w_{x}^{2} w_z }
\end{equation}
where $\kappa=w_{x}/w_{z}$ is the aspect ratio of the trapped BEC
and
\begin{equation}
f(\kappa)=\frac{1+2 \kappa^{2}}{1- \kappa^{2}}-\frac{3 \kappa^{2}
\tanh^{-1} \sqrt{1-\kappa^{2}}}{(1-\kappa^{2})^{3/2}}.
\end{equation}
Assuming that the large extension of the trap in the axial
direction means that the system is in the Thomas-Fermi limit for
that direction (i.e.\ the zero-point kinetic energy due to the
axial trapping can be ignored relative to the trap potential and
interaction energies---see \cite{odell04} for a discussion of the
Thomas-Fermi limit for dipolar gases) then the total energy
functional, scaled by the transverse trapping energy, is
approximately
\begin{equation}
\frac{H_{\mathrm{tot}}}{N \hbar \omega_{x}} \approx \frac{1}{2}
\left( \bar{w}_{x}^{2} +\frac{1}{\bar{w}_{x}^{2}} \right)+
\frac{\lambda}{4} \bar{w}_{z}^{2}-\frac{\eta}{\sqrt{2 \pi}}
\left(\varepsilon_{\mathrm{dd}}-1 \right) \frac{1}{\bar{w}_{x}^{2}
\bar{w}_{z}}
\end{equation}
where $\lambda= \omega_{z}/ \omega_{x}$ is the aspect ratio of the
trap, $\bar{w}_{x}=w_x / l_{x}$ is radial size of the BEC scaled
by the oscillator length of the trap $l_x = \sqrt{\hbar/ (m
\omega_{x})}$, and $\bar{w}_{z}=w_z / l_{z}$ is the corresponding
axial quantity. The first terms in brackets on the rhs are the
radial energy due to the trap and zero-point kinetic energy,
respectively. The next term is the axial trapping energy and the
last term is the total interaction energy, being the sum of the
$s$-wave and dipolar contributions. In order to obtain this simple
expression for $H_{\mathrm{tot}}$ we have expanded $f(\kappa)$ for
small $\kappa$ and retained only the first term. We have also used
the quantity $\eta=N a_{s}/ l_{z}$ which is a key interaction
strength parameter known from regular BECs with contact
interactions \cite{stringari96}: when $\eta$ is large the system
is in the Thomas-Fermi regime.

\begin{figure}[t]
\begin{center}
\centerline{\epsfig{figure=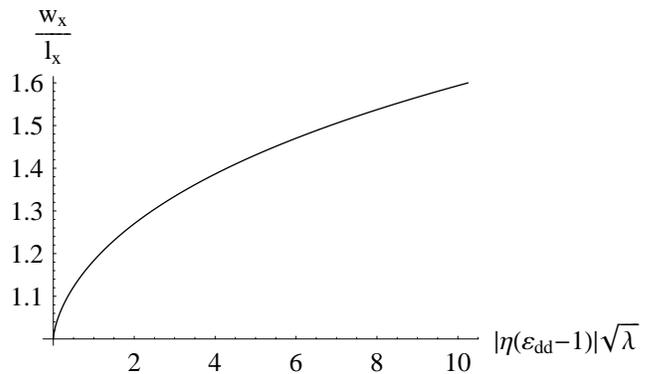, width= 8.5cm}}
\end{center}
\caption{Dependence of the radial size $w_x$ of a cigar-shaped
dipolar condensate in a 3D trap upon the system parameters.}
\label{fig:radialsize}
\end{figure}

Minimising $H_{\mathrm{tot}}$ with respect to variations of the
condensate radii gives equations for the equilibrium values of
these quantities. One finds
\begin{eqnarray}
\lambda \eta^{2} \left(\varepsilon_{\mathrm{dd}}-1 \right)^{2}=
\frac{\pi}{2} \frac{\left(\bar{w}_{x}^{4}-1
\right)^{3}}{\bar{w}_{x}^{2}} \label{eq:wx} \\
\frac{2}{\sqrt{2 \pi}} \eta \left(\varepsilon_{\mathrm{dd}}-1
\right) \frac{1}{\bar{w}_{z}}= 1- \bar{w}_{x}^{4}. \label{eq:wz}
\end{eqnarray}
Solving these equations gives the condensate radii as a function
of the various parameters. Figure \ref{fig:radialsize} illustrates
the dependence of the radius on the parameter combination $\vert
\eta (\varepsilon_{\mathrm{dd}}-1) \vert \sqrt{\lambda}$. From
Figure \ref{fig:radialsize} one sees that, all other things being
equal, the radial size increases as the dipolar interaction
strength is increased: the cigar becomes fatter. It is useful to
note that Equations (\ref{eq:wx}) and (\ref{eq:wz}) allow us to
write the total energy functional and the actual chemical
potential of the BEC in the trap (defined as
$\mu=[H_{\mathrm{kin}}+H_{\mathrm{trap}}+2H_{s}+2H_{\mathrm{dd}}]/N$)
solely in terms of the radius
\begin{eqnarray}
\frac{H_{\mathrm{tot}}}{N \hbar
\omega_{x}}=\frac{5\bar{w}_{x}^{4}-1}{4\bar{w}_{x}^{2}} \\
\frac{\mu}{N \hbar
\omega_{x}}=\frac{7\bar{w}_{x}^{4}-5}{4\bar{w}_{x}^{2}}.
\label{eq:chempot}
\end{eqnarray}

\begin{figure}[t]
\begin{center}
\centerline{\epsfig{figure=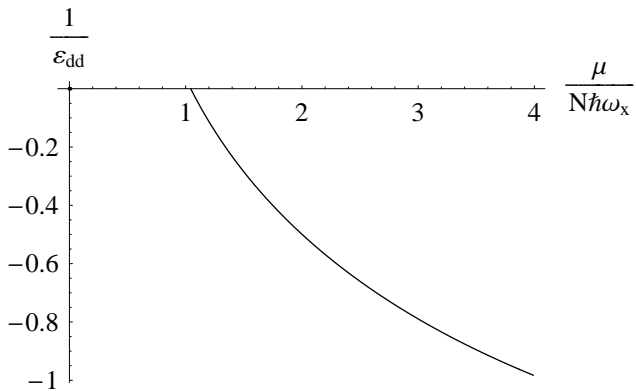, width= 8.5cm}}
\end{center}
\caption{The critical value of $1/\varepsilon_{\mathrm{dd}}$ when
the roton minimum touches the zero-energy axis plotted as a
function of the chemical potential $\mu$ in a very elongated
3D trap.} \label{fig:criticalmu1D}
\end{figure}

The solutions of equations (\ref{eq:wx}) and (\ref{eq:wz})
minimise the total energy functional in the 2D parameter space
spanned by the radii $\bar{w}_{x}$ and $\bar{w}_{z}$ for
particular fixed values of $\lambda$, $\eta$ and
$\varepsilon_{\mathrm{dd}}$. Providing the minimum is a global one
they correspond to the lowest energy solutions one can obtain from
\emph{scaling} variations. Perturbations about these equilibrium
values should result in stable oscillations which physically
correspond to monopole and quadrupole shape oscillations. If, for
new values of $\lambda$, $\eta$ and $\varepsilon_{\mathrm{dd}}$,
the minimum in this ($\bar{w}_{x},\bar{w}_{z}$) parameter space
becomes only a local minimum, or a saddle, then the system is now
either only metastable, or unstable, respectively, to scaling
perturbations. The roton instability is in principle, however, an
independent type of instability due essentially to local density
perturbations (phonons) in the same spirit as the instability we
first noted in Equation (\ref{eq:bogdispersion}) for a homogeneous
system and developed in Section \ref{sec:bog}.\ for an infinite
cylinder. Local density perturbations are not captured by the
simple ($\bar{w}_{x},\bar{w}_{z}$) parameter space and so must in
general be considered in addition to instabilities arising from
scaling perturbations. Scaling instabilities in trapped dipolar
gases have been extensively discussed before
\cite{yi,goral,santos00,giovanazzi2002b,odell04}. Our main purpose
here is to highlight the local density instability connected with
the roton minimum touching the zero energy axis.

We therefore now wish to calculate how the roton instability point
(defined by the roton minimum touching the zero-energy axis)
depends upon $\varepsilon_{\mathrm{dd}}$ and the value of the
chemical potential in a cigar-shaped 3D trap. In Figure
\ref{fig:infinitecyl} we indicated how $\varepsilon_{\mathrm{dd}}$
depends on $\bar{\mu}$ at the instability. For the case of a 3D
trap $\bar{\mu}$ is not the best quantity to consider since it
depends upon the density which is itself a non-trivial function of
$\varepsilon_{\mathrm{dd}}$ via the Equations
(\ref{eq:wx}--\ref{eq:wz}). In fact one can show
\begin{equation}
\bar{\mu}=\frac{g n(0) }{\hbar^{2}/(m
w_{r}^{2})}=\frac{4}{\sqrt{\pi}} \frac{\eta}{\bar{w}_{z}}=2
\sqrt{2} \frac{1-\bar{w}_{x}^{4}}{\varepsilon_{\mathrm{dd}}-1}.
\end{equation}
For the 3D trap we therefore prefer to illustrate the critical
value of $\varepsilon_{\mathrm{dd}}$ in terms of the true chemical
potential in the trap $\mu$, as given by Equation
(\ref{eq:chempot}). The result is shown in Figure
\ref{fig:criticalmu1D}. In the limit that
$1/\varepsilon_{\mathrm{dd}} \rightarrow 0$ we have the case of a
purely dipolar BEC with no $s$-wave interactions. We find that the
limiting value of the chemical potential at this point is $\mu/(N
\hbar \omega_{x})=1.043$ and the corresponding value of the
condensate radius is $\bar{w}_{x}=1.093$. These numbers should
only be taken as a qualitative guide rather than quantitatively
accurate since the BEC is no longer one dimensional when $\mu/(N
\hbar \omega_{x}) \approx 1$, and the effective 1D potential
approximation will begin to break down. In particular, one should
include radial excitations in order to give a consistent
treatment. This is beyond the scope of this largely illustrative
paper, but for a pancake-shaped dipolar BEC such a treatment can
be found in \cite{santos03}. Nevertheless, the very simple
treatment given here should give the general picture of what to
expect.

\section{Conclusion and outlook}
A infinite quasi-1D dipolar BEC gas would normally be unstable to
local density fluctuations if $\varepsilon_{\mathrm{dd}}>1$ on
account of the attractive nature of the dipolar interactions along
the axis of the trap. However, by reversing the sign of the
dipolar coupling using either a rotating polarizing field for
permanent dipoles \cite{giovanazzi2002b}, or a very long wavelength circularly polarized
laser beam for electrically induced dipoles, one can obtain a
stable quasi-1D system. This configuration has the possibility of
a `roton' minimum its long-wavelength axial excitation spectrum.
This minimum is tunable via parameters such as the dipolar
interaction strength and the density. As is suspected to be the
case in helium II, we speculate that the roton minimum could be
the precursor of a transition to a density wave. The tunability of
dipolar gases means that this phase transition could be explored
experimentally when quantum gases with significant dipole-dipole
interactions are realized. One of the important features of the 1D
system discussed in this paper is that the change in the order
parameter between a superfluid and a density wave does not seem to
involve any fundamental change of symmetry and so the phase
transition could be a smooth one of 2nd order. For a pancake or
fully 3D system the change in symmetry is more dramatic and these
transitions may not be smooth. We believe considerations such as
these make dipolar quantum gases very worthwhile systems to study.
We acknowledge financial support from the EPSRC (UK) and
the Marie Curie Programme of the European Commission.

\end{document}